The contribution of T2 relaxation time to diffusion MRI quantification and its clinical implications: a hypothesis.

Running title: T2 relaxation time and diffusion MRI

Yì Xiáng J. Wáng[1]*, Kai-Xuan Zhao [2], Fu-Zhao Ma [1], Ben-Heng Xiao [1]

1. Department of Imaging and Interventional Radiology, Faculty of Medicine, The Chinese University of Hong Kong, Shatin, New Territories, Hong Kong SAR, China.
2. Department of Radiology, Guangdong Provincial People's Hospital, Guangdong Academy of Medical Sciences, Southern Medical University, Guangzhou, China

* Corresponding authors:
Yì Xiáng J. Wáng, Department of Imaging and Interventional Radiology, Faculty of Medicine, The Chinese University of Hong Kong, Shatin, New Territories, Hong Kong SAR, China. E-mail: yixiang_wang@cuhk.edu.hk

**Abbreviations:**

IVIM: Intravoxel incoherent motion

$D_{slow}$ : the intravoxel incoherent motion parameter of tissue 'true' diffusion

PF: the intravoxel incoherent motion parameter of perfusion fracture

$D_{fast}$ : the intravoxel incoherent motion parameter of perfusion related diffusion.

ADC:

Eq. : Equation

Recently, we demonstrated that, using liver as the reference, IVIM derived *PF* for spleen is underestimated approximately by half [table-1, 1]. Literature also consistently reports a lower spleen $D_{slow}$ than liver $D_{slow}$ [1]. With our own data (n=20 healthy volunteers), liver $D_{slow}$ was estimated to be 1.06 ± 0.10 (×10$^{-3}$mm$^2$/s) and spleen $D_{slow}$ was 0.89 ± 0.17. Since $D_{slow}$ has limited dynamic range [2], this difference of 0.17 (×10$^{-3}$mm$^2$/s) between liver $D_{slow}$ and spleen $D_{slow}$ can be considered as substantial (this is also consistent with other reports [2]).

|  | Liver | Spleen |
|---|---|---|
| Iron mg/g [a] | 1.24±0.29 | 0.92±0.32 |
| T2* (ms) 1.5T [b] | 28.1±7.1 | 43.9±20.6 |
| T2 (ms) 1.5T [c] | 46±6 | 79±15 |
| T2 (ms) 3.0T [c] | 34±4 | 61±9 |
| T1 (ms) 1.5T [c] | 586±39 | 1,057±42 |
| T1 (ms) 3.0T [c] | 809±71 | 1,328±31 |
| flow mL/min/mL [d] |  | 1.29 ± 0.11 |
| flow mL/min/mL [e] |  | 1.4 ± 0.3 |
| flow mL/min/mL [f] |  | 1.30 ± 0.43 |
| flow mL/min/mL [g] | 1.11 ± 0.07 |  |
| flow mL/min/mL [h] | 1.47 ± 0.19 |  |
| flow mL/min/mL [i] | 1.04 |  |
| IVIM-*PF* [j] | 0.26 ± 0.077 | 0.16 ± 0.068 |
| IVIM-*PF* [k] | 0.19 ± 0.03 | 0.076 ± 0.007 |
| IVIM-*PF* [l] | 0.26 ± 0.29 | 0.09 ± 0.17 |
| Authors' data IVIM-*PF* | 0.18 ± 0.03 | 0.09 ± 0.03 |
| IVIM-$D_{slow}$ [j] | 0.92 ± 0.108 | 0.72 ± 0.085 |
| IVIM-$D_{slow}$ [k] | 1.16 ± 0.115 | 0.92 ± 0.155 |
| IVIM-$D_{slow}$ [l] | 0.88 ± 0.04 | 0.74 ± 0.29 |
| Authors' data IVIM-$D_{slow}$ | 1.06 ± 0.10 | 0.89 ± 0.17 |

**Table-1, A comparison of liver and spleen parameters (mixed male and female population).** a: Sorokin et al. Am J Hum Genet. 2022;109:1092-1104. b: Schwenzer et al. Invest Radiol. 2008;43:854-60. c. de Bazelaire et al. Radiology. 2004;230:652-9. d: Blomley et al Academic Radiology. 1997;4:13-20; e: Miles et al. Radiology. 1995;194:91-95; f: Tsushima et al. Am J Roentgenol. 1998;170:153-155; g: Blomley, et al. J Comput Assist Tomogr. 1995; 19:424-433; h, Bader et al. Invest Radiol 2000;35:539-547; i: Greenway and Stark. Physiol Rev 1971;51:23-65; j: Xu et al. J Comput Assist Tomogr. 2021;45:507-515. k: Jerome et al. J Magn Reson Imaging. 2014;39:235-40; l: Phi Van et al. Invest Radiol. 2018;53:179-185. d-h: measurement based on CT perfusion; i: based on physiological measure. Flow: input blood flow of liver from hepatic artery and poral vein and input blood flow of spleen from splenic artery. Authors' data: n=20 healthy volunteers. $D_{slow}$ unit in × 10$^3$ mm$^2$/s.

Hereby, we argue that spleen $D_{slow}$ is also substantially underestimated by IVIM if we consider liver $D_{slow}$ as the reference. Since liver and spleen have similar vessel volume and blood flow per tissue volume per minute, and the spleen is waterier than the liver (T1/T2 relaxation is 1,328/61 ms for spleen and 809/34 ms for liver at 3.0 T) [2, 3], it is quite unlikely that spleen $D_{slow}$ is much lower than liver $D_{slow}$. While there is no other reference measure for water diffusion in vivo, magnetization transfer signal ratio (MTR) measure showed a higher proportion of water molecules in the liver are bound to other macro-molecules than the water molecules in the spleen [4], which support that water molecules in the spleen have a greater extent of free diffusion. Free water molecules also allow longer T1 and T2 relaxation times than bounded water molecules. Another point of consideration is that liver has higher iron content and shorter T2* relaxation time than the spleen [3, 5], and it has been well noted that iron content and shorter T2* relaxation time are associated with lower measured ADC or $D_{slow}$ [6,7]. Despite all these, the fact that measured spleen $D_{slow}$ is much lower than liver $D_{slow}$ strongly suggests spleen $D_{slow}$ is underestimated by IVIM. Considering liver as the reference, that both fast diffusion (*PF*) and slow diffusion ($D_{slow}$) of the spleen are much underestimated is likely due to the MRI properties of the spleen such as the much longer T2 relaxation time. It is possible that longer T2 relaxation time partially mitigates the signal decay effect of various gradients on diffusion weighted image.

This phenomenon will not be limited to the spleen, thus we shall keep this in mind when we consider all MRI measured diffusion effects. Hereby we discuss two possible implications. Most liver tumors have a longer T2 relaxation time than their native normal tissue and this is considered to be associated with oedema. On the other hand, most tumors are measured with lower MRI diffusion (despite being oedematous) [8]. The reason why malignant tumors have lower diffusion value [apparent diffusion coefficient (ADC) and $D_{slow}$] are poorly understood but has been proposed to be related to a combination of higher cellularity, tissue disorganization, and increased extracellular space tortuosity [8]. These explanations may be true, but it is also possible to that many tumors have MRI properties similar to the spleen such as longer T2 (relative to the liver) and these MRI properties may also contribute to the lower MRI measured ADC and $D_{slow}$ (Fig-1). In other words, if we could hypothetically plant a piece of spleen tissue in the liver, MRI would recognize this planted spleen tissue as being similar to a tumor and measure it to have lower diffusion than the liver. Following the example of spleen tissue, *PF* of liver tumor could be underestimated by IVIM as well. Another possible implication will be the interpretation of diffusion measure after brain ischemia. After cerebral artery occlusion, diffusion decreases in the very acute phase while T2 relaxation time remains unchanged. After that, T2 relaxation time starts to increase, then diffusion starts to rise to the baseline level and then further rise to a higher level [10, 11]. If a longer T2 relaxation time 'depresses' MRI diffusion measure, at some point, such as 24 hours after the occlusion when the T2 relaxation time is already very elongated, the diffusion in the ischemic area could be highly under-estimated by MRI. However, there is a limit on how longer T2 relaxation time can 'depress' MRI diffusion measure. For example, gallbladder measures high diffusion value ($D_{slow}$ and ADC of around $3.0 \times 10^{-3}$ mm$^2$/s for our in vivo measure). Therefore, MRI measured diffusion can be complicated by many factors.

Despite being preliminary, our arguments may suggest new venues for future studies and validation for MRI derived diffusion parameters.

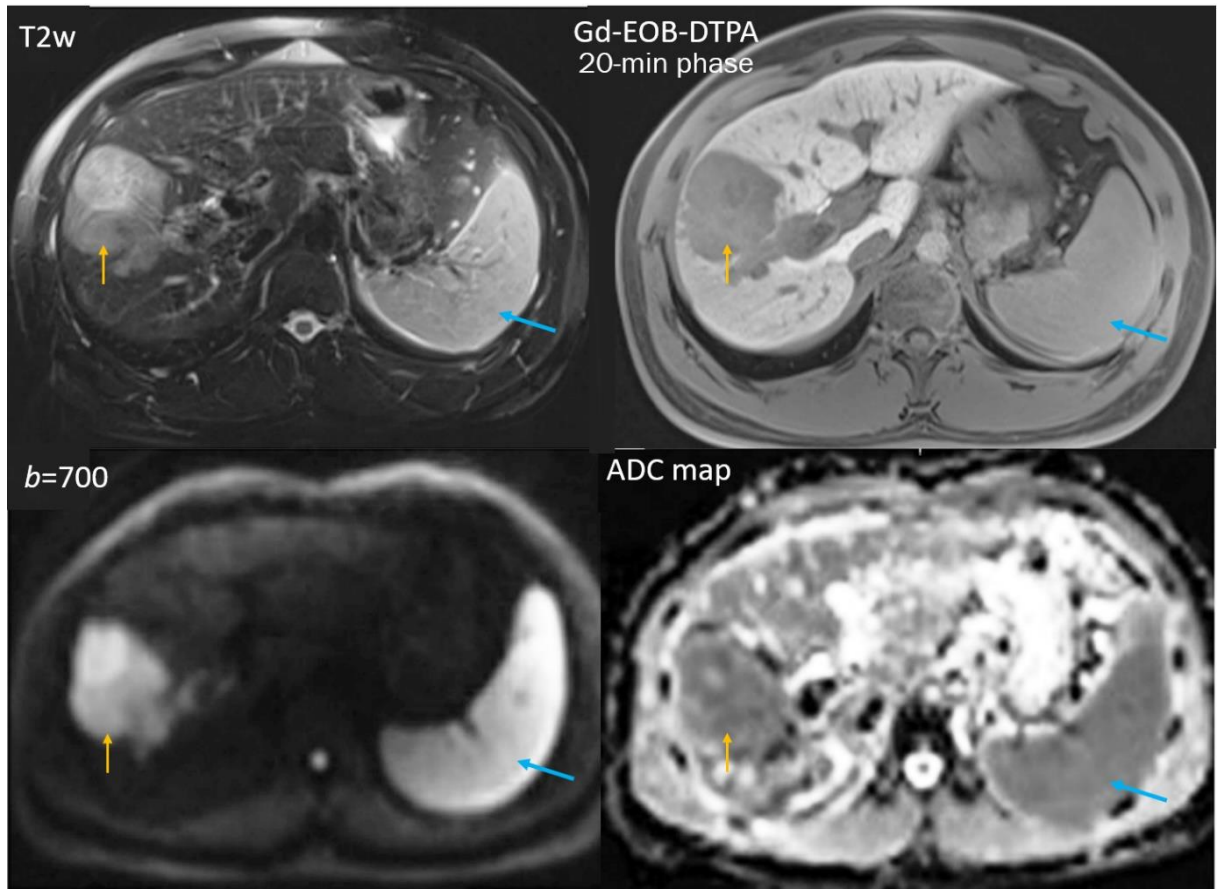

Fig-1. T2-weighted (T2w), T1-weighted Gd-EOB-DTPA enhanced 20-min hepatobiliary phase, diffusion weighted ($b$=700 s/mm$^2$) MR images, and ADC map of the liver, a hepatocellular carcinoma (orange arrow) and the spleen (blue arrow). Though the tumor has heterogeneous structures, it appears that the tumor and the spleen show many signal similarities on these four images. The lower spleen ADC value than the liver ADC value is likely due to a quantification error as described in this letter. Reproduced from Cao *at al*. [9].